\documentclass[12pt,preprint]{aastex}

\newcommand{\msun} {$M_{\odot}$}

\newcommand{\Te} {T_{\rm eff}}
\newcommand{\logg} {\log g}

\begin{document}

\title{Rare White dwarf stars with carbon atmospheres}

\author{P. Dufour\altaffilmark{1},
James Liebert\altaffilmark{1},
G. Fontaine\altaffilmark{2}, and
N. Behara\altaffilmark{3}}

\altaffiltext{1}{Steward Observatory, University of Arizona, 933 North Cherry Avenue, Tucson, AZ 85721; dufourpa@as.arizona.edu, liebert@as.arizona.edu}
\altaffiltext{2}{D\'{e}partement de Physique, Universit\'{e} 
de Montr\'{e}al, C.P. 6128, Succ. Centre-Ville, Montr\'{e}al, Qu\'{e}bec, 
Canada H3C 3J7; fontaine@astro.umontreal.ca}
\altaffiltext{3}{CIFIST, GEPI, Observatoire Paris-Meudon, 92195, France; natalie.behara@obspm.fr}

\begin{abstract}

White dwarfs represent the endpoint of stellar evolution for stars
with initial masses between approximately 0.07 \msun~and 8-10 \msun, where
\msun~is the mass of the Sun (more massive stars end their life as
either black holes or neutron stars). The theory of stellar evolution
predicts that the majority of white dwarfs have a core made of carbon
and oxygen, which itself is surrounded by a helium layer and, for
$\sim$80 per cent of known white dwarfs, by an additional hydrogen
layer$^{1-3}$. All white dwarfs therefore have been traditionally
found to belong to one of two categories: those with a hydrogen-rich
atmosphere (the DA spectral type) and those with a helium-rich
atmosphere (the non-DAs). Here we report the discovery of several
white dwarfs with atmospheres primarily composed of carbon, with
little or no trace of hydrogen or helium. Our analysis shows that the
atmospheric parameters found for these stars do not fit satisfactorily
in any of the currently known theories of post-asymptotic giant branch
evolution, although these objects might be the cooler counterpart of
the unique and extensively studied PG1159 star H1504+65 (refs
4-7). These stars, together with H1504+65, might
accordingly form a new evolutionary sequence that follow the
asymptotic giant branch.
\end{abstract}

Traces of carbon are typically observed as either neutral carbon lines
or molecular C$_2$ Swan bands (defining the DQ spectral type) in cool
helium-rich white dwarfs with effective temperatures ($\Te$) below
$\sim$ 13,000~K. The presence of carbon in the atmospheres of these
objects has been explained successfully by a model in which carbon is
dredged-up from the underlying carbon/oxygen core by the deep helium
convection zone$^8$. This model predicts a maximum contamination of
carbon at an effective temperature of $\sim$10,000~K (corresponding
approximately to the temperature at which the surface convection zone
is maximal) before gradually decreasing with lower temperature, in
agreement with atmospheric analysis determinations$^9$. We note that
although some of these objects show a very high level of carbon
pollution, helium always remains the dominant constituent of the
atmosphere (the highest carbon abundances by number (N) found are
around logN(C)/N(He) $\approx -3$). 

Atmospheric abundance determination for DQ white dwarfs has been most
successful for stars on the cool side of the maximum contamination and
very few analyses have been done for hotter objects located on the
ascending side of the curve$^{10-12}$. Recently$^{13}$, several new hot
DQ white dwarfs have been discovered in the Sloan Digital Sky Survey
(SDSS), providing a unique opportunity to test the dredge-up theory on
the ''hot side''. However, appropriate atmospheric models were not
available at that time. Because the atmospheric compositions of these
white dwarfs were expected to be helium-dominated, effective
temperature estimates (T$_{He}$) were obtained by comparing the SDSS
colours with those of pure helium composition models. Although it was
expected that such an approximation would overestimate the effective
temperature by several thousand kelvin, some of these stars were so
hot (T$_{He} \ge$ 25,000~K) that an overlap with most of the helium-rich
star (the DB spectral type) temperature range seemed unavoidable. It
was hypothesized that these stars had thinner outer helium envelopes
so that dredge-up occurred earlier in the cooling sequence. These
highly carbon-polluted white dwarfs are expected to be massive, so it
was argued that they might represent the missing high-mass tail of the
DB mass distribution$^{14}$.

Thus, it is with this scientific rationale in mind that we proceeded
with the calculation of the appropriate atmospheric models for these
objects. Because the continuum opacity of heavy elements might not be
negligible for these objects, these new models have been updated with
the latest C and O photoionization cross-sections from the Opacity
Project$^{15}$. Although the analysis of the coolest objects ($\Te \le
$15,000~K) is straightforward (results will be presented elsewhere;
manuscript in preparation), we quickly realized that no combination of
carbon and helium could successfully reproduce the observed features
(mostly C\textsc{ii} lines) in the optical spectrum of the hottest ones by
assuming a helium-dominated atmosphere. Indeed, such models predict
the presence of a strong He\textsc{i} $\lambda$=4,471 line that is not observed
spectroscopically in our sample of hot DQ stars. We thus concluded
that a good fit to both the spectra and the energy distribution was
possible only by considering atmospheres made primarily of carbon,
with little or no trace of hydrogen and helium.  

Figure 1 shows our fits to the optical spectrum and photometric energy
distribution of one such star using a grid of pure carbon atmosphere
models, as well as models containing traces of hydrogen and
helium. The fitting method is similar to that used for cooler DQ and
needs not to be repeated here$^9$. This particular case shows a small
trace of hydrogen which we believe is more likely to be the result of
accretion from the interstellar medium than of primordial origin. We
found eight more similar objects in the SDSS white dwarf
catalogue$^{16}$, all of which were found to have a carbon-dominated
atmosphere and a temperature between 18,000 and 23,000~K, although for
some faint stars, detailed analysis is more uncertain because of the
poor signal-to-noise ratio of the observations. Only crude upper
limits for hydrogen and helium (of the order of logN(C)/N(H) $\ge$ 1.5 and
logN(C)/N(He) $\ge$ 1.5) can be obtained from the other spectra. Our
detailed model atmosphere fits provide fair estimates of the effective
temperature and chemical composition, but higher-quality data (which
we hope to obtain in December 2007, weather permitting) are essential
for a precise measure of the surface gravity (and thus mass) of these
objects.

These new stars are too hot to be explained by the standard convective
dredge-up scenario and so we sought another explanation. Natural
progenitor candidates to consider are the hot PG1159 stars. The
latter are probably the result of a late He-shell flash$^{17}$ (the
so-called ''born-again'' scenario) at the end of the
post-asymptotic-giant-branch phase, which almost completely eliminates
the remaining hydrogen and mixes material from the interior with the
helium envelope. As a result, these objects re-enter the white dwarf
cooling track, but this time they have a surface composition that is a
mixture of helium, carbon, oxygen, and little or no hydrogen ($\Te$
between $\sim$75,000 and 200,000 K; typical abundances$^{18}$, in mass
fraction, of He, C, O and Ne are 33$\%$, 50$\%$, 15$\%$ and 2$\%$). However, as
these stars cool down, gravitational diffusion rapidly separates the
heavier elements from the helium that tends to float to the surface
and by the time they have reached 25,000 K, helium completely
dominates the surface composition.

Models exploring the evolution of stars 9-11 \msun~have produced white
dwarfs with O-Ne-Mg cores and CO envelopes and possibly little He and
H; such models may explain our peculiar stars$^{19-21}$. However, it
is unclear what quantity of He and H should remain for such
stars. Furthermore, the resulting white dwarfs have high surface
gravities ($\logg \approx 9$, in centimetre-gram-second units) and
would thus show extremely broad carbon lines that are incompatible
with the observed widths of the lines (Fig. 1 shows that $\logg
\approx 8$ is more likely, although analysis of a better spectrum is
needed for more precision). The most likely scenario to explain the
very existence of hot DQ white dwarfs with carbon-dominated
atmospheres is that these stars are the progenies of objects such as
H1504+65. The latter is a unique object among the known PG1159 stars
so far. It is the hottest specimen of its class at $\Te \approx$
200,000 K, and its atmospheric composition very unusual, with a mass
fraction of $\sim 50\%$ C and $\sim 50\%$ O plus small traces of heavier
elements, but no detectable helium or hydrogen. It is currently
believed that H1504+65 is essentially a bare stellar nucleus produced
by a particularly violent post-asymptotic-giant-branch very late
thermal pulse that has destroyed in large part the remaining stellar
envelope containing helium and hydrogen. The best available
simulations of such an event, those of ref. 17, suggest indeed that,
following the last He flash, the full envelope becomes convective and
extends to deep enough layers for H and He to be completely consumed,
except that very small traces of He may survive.

The uniformity of the chemical composition of the envelope of H1504+65
is believed to be maintained for a while by a residual wind but, with
time, diffusion becomes dominant and He, C and O must separate under
the influence of gravitational settling. The idea is that the wind
slowly dies out with cooling (decreasing luminosity) and that the
braking effect it provides on element separation becomes less and less
efficient. Hence, the small residual trace of helium believed to exist
in the envelope of H1504+65 eventually diffuses upward to form a thin
layer above a C-enriched and O-depleted mantle. The total mass of an
atmosphere is tiny ($\sim 10^{-14}$ to $10^{-15}$ M$_*$, where M$_*$
is the mass of the star), so there is ultimately enough accumulated He
to form a full atmosphere and the descendant of H1504+65 would now be
''disguised'' as a He-atmosphere white dwarf after the PG1159 evolutionary
phase. With further cooling, a convection zone develops in the
C-enriched mantle owing to the recombination of that element while the
overlaying He layer remains radiative. At some $\Te$, whose exact
value is not yet known due to a current lack of proper models, it is
conjectured that the subphotospheric C convection zone becomes active
enough to be able to dilute from below the overlaying He layer. At
that point the star would undergo a dramatic spectral change,
transforming itself from a He-dominated atmosphere white dwarf to a
star with a C-dominated atmosphere, because the mass in the C
convection zone is orders of magnitude larger than the mass of the He
layer.

Hence, the former PG1159 star H1504+65, showing initially a mixed C
and O atmosphere, would now show a C-dominated atmosphere after an
intermediate phase in which it would have been observed as a
He-atmosphere white dwarf. We note, however, that helium must
ultimately reappear at the surface in enough quantity to form a
helium-rich atmosphere because no carbon-rich object has ever been
discovered at lower temperatures ($\Te \le$ 15,000 K). The remaining
helium probably floats again at the surface when the star cools down,
perhaps turning the carbon-rich objects into DQ stars belonging to the
second sequence of high carbon abundances$^{11,22}$. We are at present
making evolutionary calculations to test the above scenario.

To conclude, we roughly estimate that the space density of carbon-rich
white dwarfs, assuming a 50$\%$ completeness for the targeting in
SDSS$^{16}$, is between $\sim 2.1\times10^{-6}$~pc$^{-3}$ and $\sim
7.0\times10^{-8}$~pc$^{-3}$ depending on the preferred value of
$\logg$ (and thus radius) and $\Te$ for the stars. For comparison, the
local space density of white dwarfs$^{23}$ is estimated to be $\sim
5.0\times10^{-3}$~pc$^{-3}$. White dwarfs with carbon dominated
atmospheres are thus intrinsically rare and a relatively large volume
of space (as was done by the Sloan Digital Sky Survey) had to be
surveyed before many of them could be found.

\noindent 1. Iben, I., Jr. On the frequency of planetary nebula nuclei powered by helium burning and on the frequency of white dwarfs with hydrogen-deficient atmospheres. Astrophys. J.  {\bf277}, 333-354 (1984).\\
2. Koester, D. $\&$ Schoenberner, D. Evolution of white dwarfs. Astron. Astrophys.  {\bf154}, 125-134 (1986).\\
3. D'Antona, F. $\&$ Mazzitelli, I. Evolutionary times of white dwarfs-Long or short? IAU Colloq.  {\bf95}, 635-637 (1987).\\
4. Nousek, J. A. et al. H1504+65-an extraordinarily hot compact star devoid of hydrogen and helium. Astrophys. J.  {\bf309}, 230-240 (1986).\\
5. Werner, K. NLTE analysis of the unique pre-white dwarf H 1504 + 65. Astron. Astrophys.  {\bf251}, 147-160 (1991).\\
6. Werner, K. $\&$ Wolff, B. The EUV spectrum of the unique bare stellar core H1504+65. Astron. Astrophys.  {\bf347}, L9-L13 (1999).\\
7. Werner, K., Rauch, T., Barstow, M. A. $\&$ Kruk, J. W. Chandra and FUSE spectroscopy of the hot bare stellar core H 1504+65. Astron. Astrophys.  {\bf421}, 1169-1183 (2004).\\
8. Pelletier, C., Fontaine, G., Wesemael, F., Michaud, G. $\&$ Wegner, G. Carbon pollution in helium-rich white dwarf atmospheres. Time-dependent calculations of the dredge-up process. Astrophys. J.  {\bf307}, 242-252 (1986).\\
9. Dufour, P., Bergeron, P. $\&$ Fontaine, G. Detailed spectroscopic and photometric analysis of DQ white dwarfs. Astrophys. J.  {\bf627}, 404-417 (2005).\\
10.  Wegner, G. $\&$ Koester, D. Atmospheric analysis of the carbon white dwarf G227-5. Astrophys. J.  {\bf288}, 746-750 (1985).\\
11. Thejll, P., Shipman, H. L., MacDonald, J. $\&$ Macfarland, W. M. An atmospheric analysis of the carbon-rich white dwarf G35 - 26. Astrophys. J.  {\bf361}, 197-206 (1990).\\
12. Desharnais, S., Wesemael, F., Chayer, P. $\&$ Kruk, J. W. FUSE observation of cool DB white dwarfs. ASP Conf. Ser.  {\bf372}, 265-268 (2007).\\
13. Liebert, J. et al. SDSS white dwarfs with spectra showing atomic oxygen and/or carbon lines. Astron. J.  {\bf126}, 2521-2528 (2003).\\
14. Beauchamp, A. D\'etermination des Param\`etres Atmosph\'eriques des \'etoiles Naines Blanches de Type DB. PhD thesis, Montr\'eal (1995).\\
15. Behara, N. $\&$ Jeffery, C. S. LTE model atmosphere with new opacities. 1. Methods and general properties. Astron. Astrophys.  {\bf451}, 643-650 (2006).\\
16. Eisenstein, D. J. et al. A catalog of spectroscopically confirmed white dwarfs from the Sloan Digital Sky Survey data release 4. Astrophys. J.  {\bf167} (Suppl.), 40-58 (2006).\\
17. Herwig, F., Bl\"ocker, T., Langer, N. $\&$ Driebe, T. On the formation of hydrogen-deficient post-AGB stars. Astron. Astrophys.  {\bf349}, L5-L8 (1999).\\
18. Werner, K. $\&$ Herwig, F. The elemental abundances in bare planetary nebula central stars and the shell burning in AGB stars. Publ. Astron. Soc. Pacif.  {\bf118}, 183-204 (2006).\\
19. Garcia-Berro, E. $\&$ Iben, I. On the formation and evolution of super-asymptotic giant branch stars with cores processed by carbon burning. 1: SPICA to Antares. Astrophys. J.  {\bf434}, 306-318 (1994).\\
20. Garcia-Berro, E., Ritossa, C. $\&$ Iben, I. On the evolution of stars that form electron-degenerate cores processed by carbon burning. III. The inward propagation of a carbon-burning flame and other properties of a 9 \msun~model star. Astrophys. J.  {\bf485}, 765-784 (1997).\\
21. Ritossa, C., Garcia-Berro, E. $\&$ Iben, I. On the evolution of stars that form electron-degenerate cores processed by carbon burning. V. Shell convection sustained by helium burning, transient neon burning, dredge-out, URCA cooling, and other properties of an 11 \msun~population I model star. Astrophys. J.  {\bf515}, 381-397 (1999).\\
22. Koester, D. $\&$ Knist, S. New DQ white dwarfs in the Sloan Digital Sky Survey DR4: confirmation of two sequences. Astron. Astrophys.  {\bf454}, 951-956 (2006).\\
23. Holberg, J. B., Oswalt, T. D. $\&$ Sion, E. M. A determination of the local density of white dwarf stars. Astrophys. J.  {\bf571}, 512-518 (2002).\\

\acknowledgements This work has been partially supported by the NSF
for work on SDSS white dwarfs. This work was also supported in part by
the NSERC (Canada).

{\bf Author Information} Reprints and permissions information is
available at www.nature.com/reprints. Correspondence and requests for
materials should be addressed to P.D. (dufourpa@as.arizona.edu).

\clearpage

\clearpage
\begin{figure}[p]
\figcaption[f1.eps] {{\bf Fit to the optical spectra and energy
distribution for a carbon-rich white dwarf. a}, The thick red line
represents our $\logg$ = 8 best fit (the parameters are indicated in
the panel) while the blue line represents a solution for $\logg$ =
9. Hydrogen abundance is determined by fitting $H\beta$
($\lambda$=4,861$\rm \AA$) while helium abundance is constrained by
the absence of the He\textsc{ii} $\lambda$ =4,471$\rm \AA$ line. {\bf
b}, Photometric measurements in the u, g, r, i and z bands are
represented by error bars, while the average model fluxes from the
same best $\logg$ = 8 model are shown by filled circles. Both {\bf a}
and {\bf b} have the same y axis.
\label{fg:f1}}
\plotone{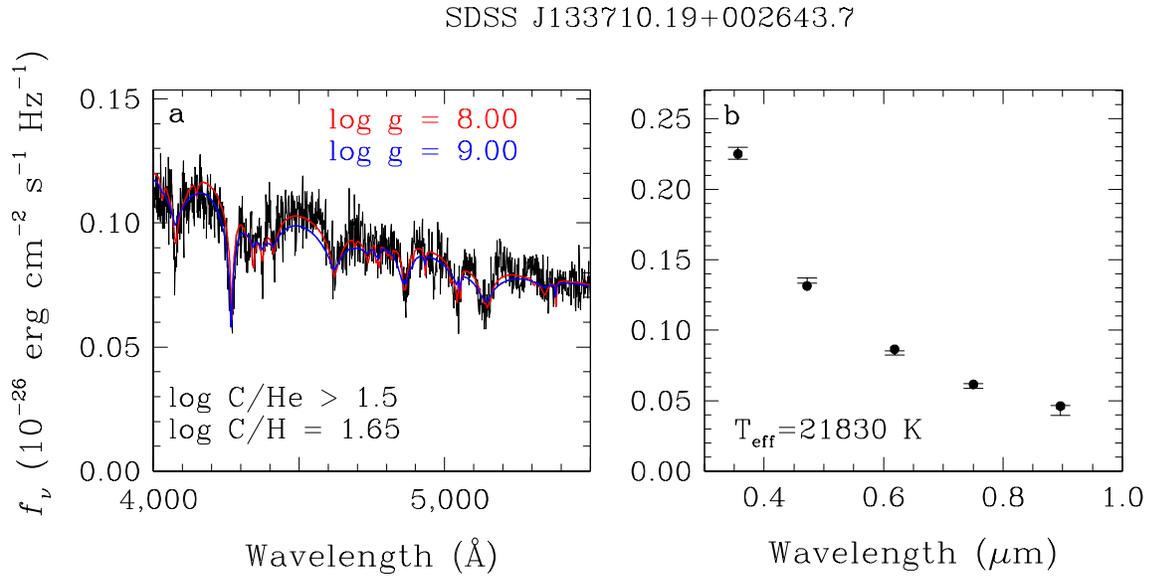}
\begin{flushright}
Figure \ref{fg:f1}
\end{flushright}
\end{figure}

\end{document}